\documentclass[12pt,a4paper,reqno]{amsart}
\usepackage[dvips]{graphicx}
\usepackage{mathrsfs}
\usepackage{amsthm}
\usepackage{amstext}
\usepackage{amsmath}
\usepackage{amsfonts}
\newcommand{\eu}{{\rm e}}
\newcommand{\dd}{{\rm d}}

\begin{document}

\title[Orthogonal polynomial method and odd vertices]{Orthogonal polynomial method and odd vertices in matrix models}
\date{}
\maketitle
\begin{center}
Ettore Minguzzi \footnote{e-mail: eminguzz@mailbox.difi.unipi.it}\\
{\small \it Dipartimento di Fisica dell'Universit\`a, Pisa 56100, Italy and \\ 
 INFN, Sezione di Pisa}
\end{center}

\begin{abstract} 
We show how to use the method of orthogonal polynomials for integrating, in the planar approximation, the partition function of  one-matrix models with a potential    with even or odd vertices, or any combination of them.
\end{abstract}

\section{Introduction}
 The method of orthogonal polynomials is a powerful technique for the non perturbative integration of matrix models over one~\cite{Be} or  more matrices~\cite{M1} in  particular with even potential, \emph{i.e.} with vertices with an even number of legs. Indeed, with even potential,  the  calculation simplifies both because the integrals are well defined and, as we shall see, the number of equations needed to solve the problem is smaller.
On the other hand the model with odd vertices, in particular with cubic vertices is more natural in a number of problems; \emph{e.g.} in the dynamical triangulation model of quantum gravity, where the random surface is given by a polyhedron with triangular faces, the order of the vertices appearing in the dual graphs is always three. Br\'ezin et~al.~\cite{Br} solved the problem with cubic vertices using the  saddle point technique. Bessis \cite{Be1} introduced an alternative method (the orthogonal polynomial method) which to some extent appears more powerful \emph{e.g.} in dealing with matrix model with more than one matrix variable~\cite{M1}. In particular, the orthogonal polynomial method has been proved useful in the treatment of a cubic vertex  two-matrix model~\cite{BK} in the context of the Ising model on a random planar lattice.

The purpose of this paper  is to show, in a systematic way, how to extend the orthogonal polynomial method to arbitrary vertices, both even and odd and any combination of them.  We shall follow the article of Bessis \emph{et al.} \cite{Be} generalizing some aspects to the case of odd vertices, in particular we shall recover, for the simplest case of cubic vertices, the result of \cite{Br} for spherical topology. Hopefully such a treatment can be extended to higher genus.

The use of mixed vertices \emph{e.g.} cubic plus quartic vertex, allows us to write a well defined \emph{i.e.} convergent, partition function by adding to the cubic interaction  a quartic term which makes the action bounded from  below, and thus the integral giving the partition function well defined. At the end one can take the limit when the coupling constant of the  quartic vertex goes to zero.

\begin{figure}[!ht]
\begin{center}
\includegraphics[angle=0, width=9.2cm]{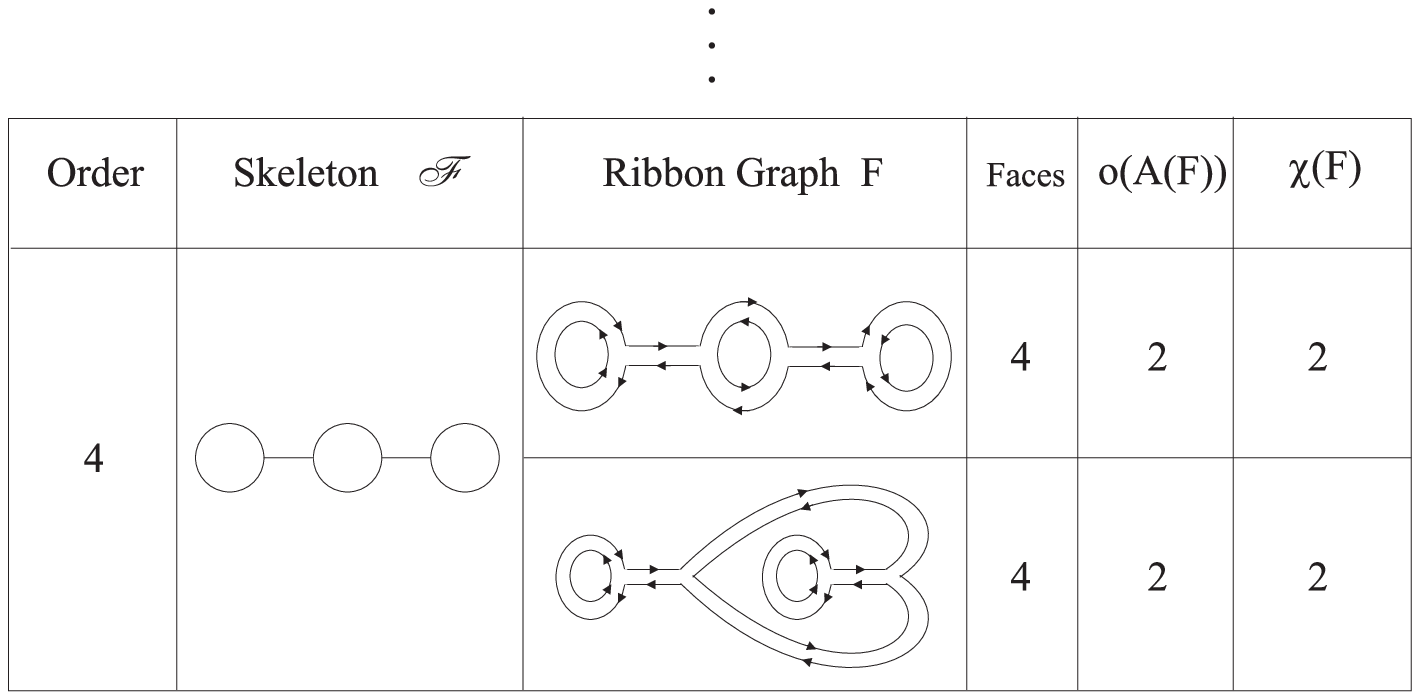}
\end{center}
\label{dis1}
\end{figure}

\begin{figure}[!hb]
\begin{center}
\includegraphics[angle=0, width=9.2cm ]{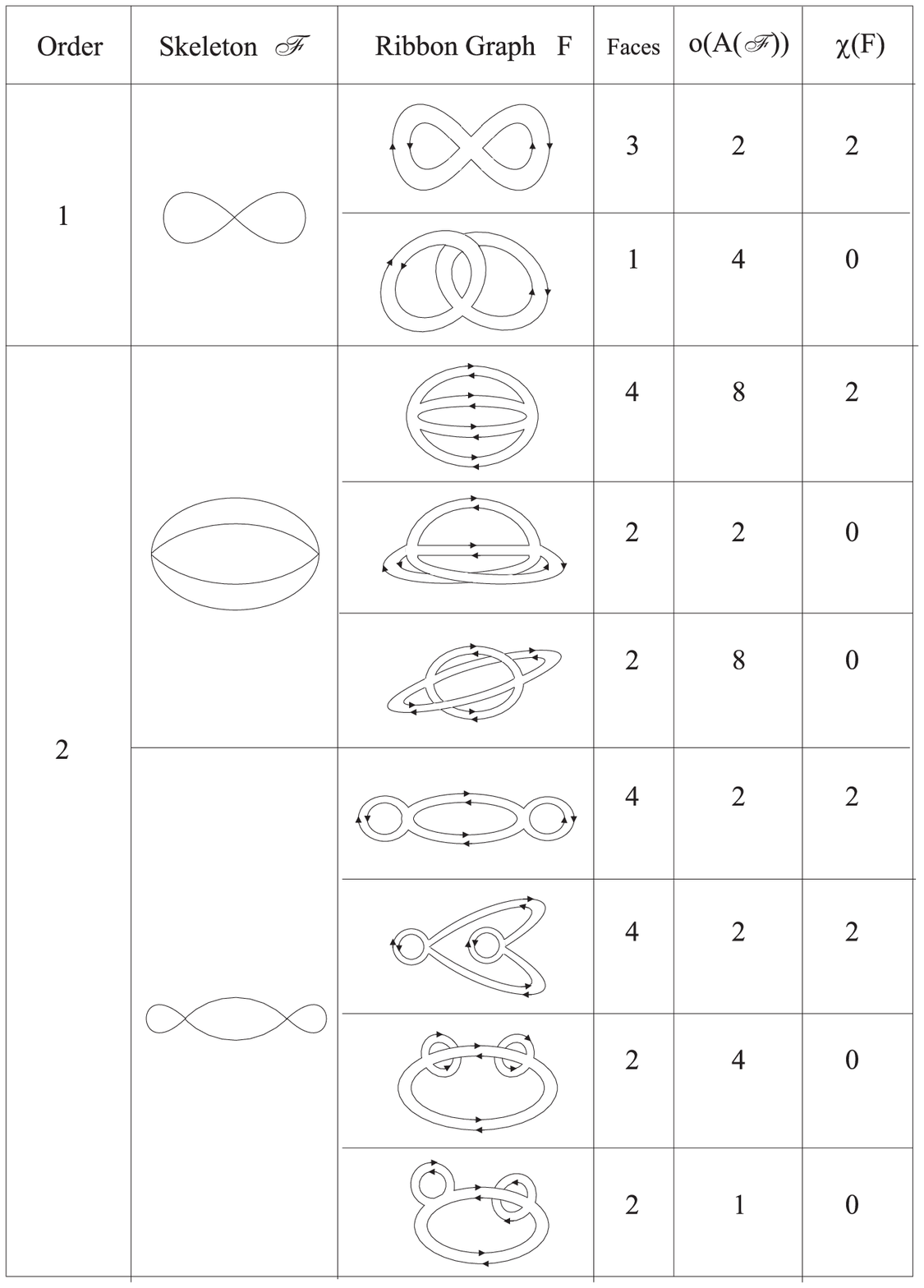}
\end{center}
\caption{First and second order connected graphs with quartic vertex.}
\label{dis2}
\end{figure}

 We start from the partition function
\begin{equation} \label{part}
Z_N({\bf g})= \int \dd M\, \eu^{-{\rm tr}S(M)}
\end{equation}
where the integration is over an hermitian matrix of order $N$ and where the action is given in general by
\begin{eqnarray}
-S(M) & = & -\frac{1}{2}M^2+V(M)  \\
V(M) & = & \sum_{j=3}^k \frac{g_j}{jN^{\frac{j}{2}-1}} M^j \qquad {\rm with} \ k \ {\rm even \ and} \ {\rm Re}(g_k)<0 \ .
\end{eqnarray}
In the following we shall use also the variables $\bar{g}_i=\frac{g_i}{N^{\frac{i}{2}-1}}$.
As we can see the potential is composed of a combination of vertices, however, in order  it to be bounded from above, and  thus the integral  be well defined, it is needed that the vertex with highest valence be even, and that the real part of the related coupling constant be negative. We are interested in the functions $e_h({\bf g})$
\begin{equation}
 \ln \frac{Z_N({\bf g})}{Z_N({\bf 0})}=\sum_{h=0}^{\infty}N^{2-2h} e_h({\bf g}) 
\end{equation}
where $2-2h$ is the Euler characteristic of the oriented ribbon graphs to be summed in the perturbative expansion of the functions $e_h({\bf g})$. Indeed, denoting such graphs with capital letters, each function $e_h$ admits the following expansion \cite{Be} 
\begin{equation} \label{top}
 e_h({\bf g})=\sum_{\substack{G \ connected \ of \\ genus \ h}} \frac{\prod_ig_i^{v_i(G)}}{o(A(G))} 
\end{equation}
where $v_i(G)$ is the number of vertices with valence $i$ of the ribbon graph $G$ and  $o(A(G))$ is the order of the group of automorphisms of the graph \footnote{An automorphism of an  oriented ribbon graph is defined in the following way. First let us identify the oriented ribbon graph as a common graph plus a cyclic ordering on the sets of half-edges attached to each vertex and then define an automorphism of the oriented ribbon graph as an automorphism of the graph which leaves the ordering of each vertex unchanged. It's clear that the automorphism must send each vertex into a vertex of equal valence.}. Figures \ref{dis1}, \ref{dis2} show the simplest connected ribbon graphs in the cubic and quartic case.  
Models with odd vertices will be regularized as explained above \emph{i.e.} by adding a regulating even interaction which at the end is put to zero.  We shall see explicitly how this method works for the cubic vertex.
\begin{figure}[!hb]
\caption{Second and fourth order connected graphs  with cubic vertex.}
\begin{center}
\includegraphics[angle=0, width=10cm]{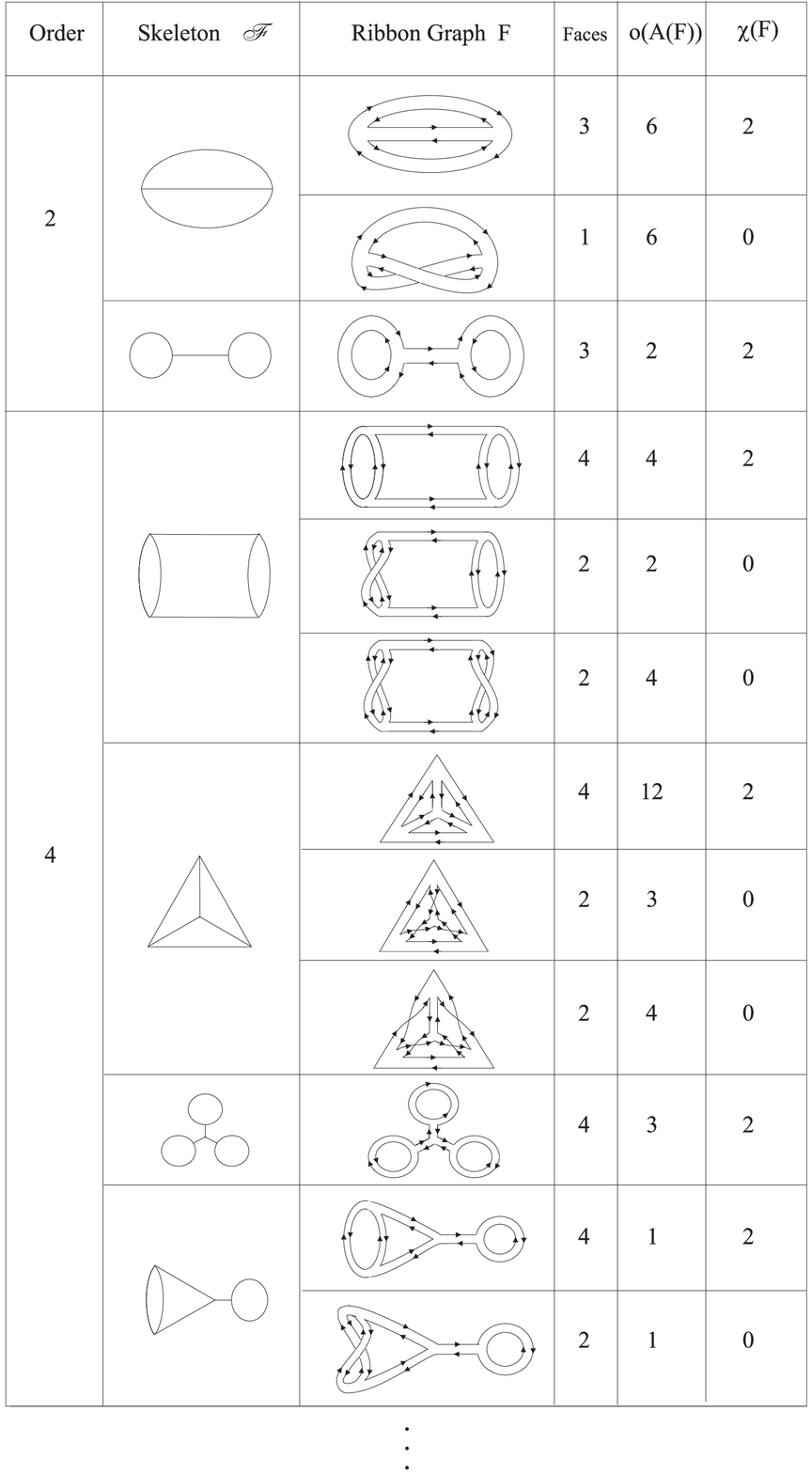}
\end{center}
\end{figure}
\clearpage

\section{The method of orthogonal polynomials}

A change of integration variables in (\ref{part}) leads us to the integration over  the eigenvalues $\lambda_i$ of the diagonal matrix $\boldsymbol{\lambda}$
\begin{equation} \label{diag}
Z_N({\bf g})= \int \dd M\, \eu^{-{\rm tr}S(M)}= k_H \int \prod_i \dd \lambda_i \,\Delta^2(\boldsymbol{\lambda}) \eu^{-\sum_i S(\lambda_i)} 
\end{equation}
where $\Delta(\boldsymbol{\lambda})=\prod_{\alpha<\beta} (\lambda_{\beta}-\lambda_{\alpha})$ is the Vandermonde determinant.
We obtain the value of the constant $k_H$  using the results in \cite{M2}: $k_H=\frac{\pi^{\frac{N^2-N}{2}}}{\prod_{j=1}{N} j!}$.
As we see the argument of the integral is the product of the Vandermonde determinant squared and a factorizable function of the eigenvalues, this feature makes the orthogonal polynomial method applicable. Let us introduce the measure $\dd \mu (\lambda)= \dd \lambda \,\eu^{- S(\lambda)}$, and the orthogonal polynomials $P_n(\lambda)$
\begin{equation}
 \int^{+ \infty}_{- \infty}\dd \mu (\lambda)\,P_n(\lambda)P_m(\lambda)=h_n \delta_{nm} \end{equation}
where $P_n(\lambda)$ is normalized by the condition that the coefficient of the  term with highest degree equals $1$
\begin{equation}
P_n(\lambda)=\lambda^n+ \ldots \ .
\end{equation}
The polynomials $P_n(\lambda)$ can be obtained in a constructing way \emph{e.g.} by the  Gram-Schmidt orthogonalization procedure from the monomials  $1$, $\lambda$, $\lambda^2$,  $\ldots$. A simple analysis of this procedure shows that the polynomials $P_j$ have the well defined parity $(-1)^j$ if the action $S(\lambda)$ is even.  Every polynomial of degree $n$ can be rewritten as a linear combination of $P_m$ with $m \leq n$. The Vandermonde determinant in (\ref{diag}) can be rewritten as
\begin{equation}
 \Delta={\rm det} \parallel \lambda_i^{j-1} \parallel={\rm det} \parallel P_{j-1}(\lambda_i) \parallel=\sum_{\sigma}(-1)^{p(\sigma)}\prod_i^NP_{\sigma(i)-1}(\lambda_i) 
\end{equation}
where the second equality is due to the fact that adding to a column a linear combination of the other columns does not change the determinant of the matrix;  $(-1)^{p(\sigma)}$ stands for the sign of the permutation $\sigma$. We can take advantage of the coupling of the orthogonal polynomials due  $\Delta^2$ in (\ref{diag}) to obtain the partition function in terms of the norm of the orthogonal polynomials
\begin{equation} \begin{split}
 Z_N({\bf g})  & =  k_H \! \sum_{\sigma_1, \ \sigma_2}(-1)^{p(\sigma_1)}(-1)^{p(\sigma_2)} \prod_i^N \int \! \dd \mu(\lambda_i)\,P_{\sigma_1(i)-1}(\lambda_i)P_{\sigma_2(i)-1}(\lambda_i)  \\
{}  & =   k_H \! \sum_{\sigma_1, \ \sigma_2} \! (-1)^{p(\sigma_1)}(-1)^{p(\sigma_2)} \delta_{\sigma_1\sigma_2} \prod_i h_{\sigma_1(i)-1}=k_H N!\, h_0h_1 \ldots h_{N-1} \ .
\end{split} \end{equation}
Let us rewrite this expression in a different form.
The following equation is valid
\begin{equation}
\lambda P_n(\lambda)=P_{n+1}(\lambda)+A_n P(\lambda)+R_n P_{n-1}(\lambda) 
\end{equation}
where the terms with index less than $n-1$ are absent because after moltiplication by $\lambda$ they do not reach the degree $n$ and thus are orthogonal to $\lambda P_n$. For parity reasons when $S(\lambda)$ is even $A_n$ vanishes. We shall refer to the preceding equation as the {\em step equation} because its repeated application enables us to calculate $\lambda^iP_n(\lambda)$ using an analogy with all possible staircases $i$ steps long. This method will be developed in the following section. Since
\begin{equation} \begin{split}
 h_{n+1}  & =   \int \! \! \dd \mu(\lambda) \,P_{n+1}\lambda P_n(\lambda )  \\
{} & =   \int \! \! \dd \mu(\lambda) \,[P_{n+2}(\lambda) + \! A_{n+1}P_{n+1}(\lambda)+ \!R_{n+1}P_n(\lambda)] P_n(\lambda)=R_{n+1}h_n
\raisetag{42pt}
\end{split} \end{equation}
the partition function can be rewritten as
\begin{equation} 
Z_N({\bf g})=k_H N!\, h_0^N R_1^{N-1}\ldots R_{N-2}^2 R_{N-1} 
\end{equation}
where $h_0=\int \dd \lambda \eu^{-S(\lambda)}$. Before passing to the limit for large $N$, we must compute
\begin{equation} \label{EN}
E_N({\bf g}) \!  = \! \frac{1}{N^2}\ln \frac{Z_N({\bf g})}{Z_N({\bf 0})}=\frac{1}{N}\sum_{n=1}^{N} \! \left( \! 1- \! \frac{n}{N} \right) \ln \! \frac{R_n({\bf g})}{R_n({\bf 0})}+\! \frac{1}{N} \ln \frac{h_0({\bf g})}{h_0({\bf 0})} \ . 
\end{equation} 
We can  prove that the last term on the r.h.s.~of (\ref{EN}) is negligible using the same perturbative method used for the partition function. The main difference is that $h_0({\bf g})$ is integrated over a scalar variable whereas  partition functions such as $Z_N({\bf g})$ are integrated over a matrix variable. Expanding the action in $h_0$  and  recalling the proof of the topological expansion \cite{Be}, we notice the absence of the typical contribution due to the propagator deltas, namely a factor $N$ for each face. Finally  we have to add a factor $N^{-F}$ which yields
\begin{equation}
\ln \frac{h_0({\bf g})}{h_0({\bf 0})}=\sum_{G \ connected} \frac{N^{\chi(G)-F}\prod_ig_i^{V_i}}{o(A(G))}  \ .
\end{equation}
 Since $\chi(G)-F= -\sum_{i=3}V_i(\frac{i}{2}-1) \leq -1 $, we have
\begin{equation}
 \frac{1}{N}\ln \frac{h_0({\bf g})}{h_0({\bf 0})}=O(N^{-2}) 
\end{equation}
which vanishes for large N. Thus
\begin{equation}
e_0({\bf g}) = \lim_{N \to \infty}\,E_N({\bf g})  = \lim_{N \to \infty} \,\frac{1}{N}\sum_{n=1}^{N} \left(1-\frac{n}{N} \right) \ln \frac{R_n({\bf g})}{R_n({\bf 0})} 
\end{equation}
\section{The number of staircases}
\begin{figure}[ht]
\begin{center}
\includegraphics[angle=0, width=9.5cm]{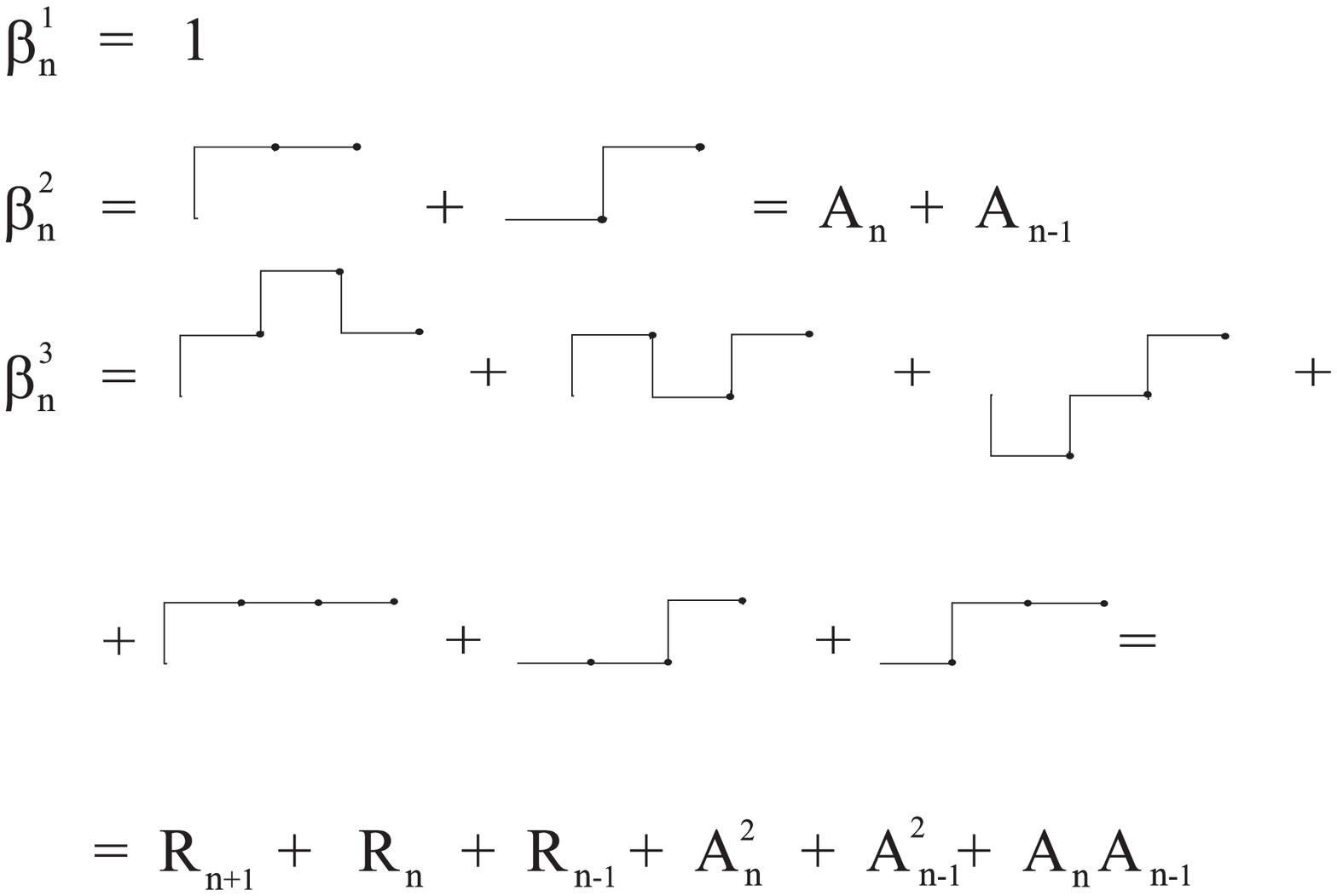}
\end{center}
\caption{$\beta^i_n$ computed from the staircases.}
\label{gradini}
\end{figure}

We shall need, in the following,  the quantities  $\beta_n^i$ defined by
\begin{equation}
h_n \beta_n^i = \int \dd \mu(\lambda)\,P_n(\lambda) \lambda^i P_{n-1}(\lambda) \ .
\end{equation}
We   devote the present section to the calculation of the above integral. To compute $\lambda^i P_{n-1}$ we take advantage  of an analogy with all staircases of $i$ steps; where each step can go up, come down, or stay at the same level. The analogy comes from a repeated application of the step equation. After the integration only the staircases which end one step up, contribute. Each of them represents a product of factors: if a step is down from  level $n$ to the level $n-1$ we add a factor $R_n$, and if it stays at the same level $n$ we add a factor $A_n$. Figure~\ref{gradini} shows an example of this kind of calculation.
Since every coefficient $A_j, \ R_j$, is a function of the index $j$ it would be difficult to hand the final expression for $\beta_n^i$; luckily, as we shall see, the  planar limit ( $N \to \infty$ ) will enable us to neglect the differences among these quantities relative to different levels. In this limit we must compute the expression for  $\beta_n^i$ supposing that each step down yields a factor $R$, and each step that stays at the same level yields a factor $A$. Thus the question is: How many are the staircases of $i$ steps whose final effect is to go up one step? Let  $j$ be the steps of type A, then the other $i-j$ are divided in $p$ steps up and $p-1$ steps down so that  $i=j+2p-1$. Without the $A$ steps there are  ${2p-1 \choose p}$ staircases of $2p-1$ steps whose final effect is to go up one step. Inside these staircases we want to insert the remaining  $j$ levels of type $A$: there are  $2p$ places where they can be inserted, and, for a fixed staircase, there are  ${2p+j-1 \choose j}$ choices. Finally the number of staircases of $i$ steps whose final effect is to go up one step is
\begin{equation}
\sum_{p=1}^{[ \frac{i+1}{2}]} \frac{i!}{(i-2p+1)!p!(p-1)!} \ ,
\end{equation}
where $[ \: ]$ stands for the integer part, and,  denoting by $\tilde{\beta}^i$ the continuum value of $\beta^i_n$, we have  
\begin{equation} \label{beta}
 \tilde{\beta}^i=\sum_{p=1}^{[ \frac{i+1}{2} ]} \frac{i!}{(i-2p+1)!p!(p-1)!}\,A^{i-2p+1}R^{p-1}
\end{equation}
where the tilde reminds the replacement $A_j \to A, \ R_j \to R$. The values of $\tilde{\beta}^i$ for the first few $i$ are
\begin{equation} \begin{split}
\tilde{\beta}^2 &= 2A \\
\tilde{\beta}^3 &= 3A^2+3R \\
\tilde{\beta}^4 &= 4A^3+12AR \ .
\end{split} \end{equation}
Analogously we define
\begin{equation}
h_{n+1} \gamma_n^i = \int \dd \mu(\lambda)\,P_{n+1}(\lambda) \lambda^i P_{n-1}(\lambda) \ .
\end{equation}
By the same technique used for $\beta^i_n$ we find
\begin{equation} 
\tilde{\gamma}^i=\sum_{p=2}^{[ \frac{i}{2}+1 ]} \frac{i!}{(i-2p+2)!p!(p-2)!}\,A^{i-2p+2}R^{p-2} \ .
\end{equation}
Finally we define
\begin{equation} \label{delta}
 \tilde{\delta}^i= \! \tilde{\beta}^i\!-R\tilde{\gamma}^{i-1}\!\!-A\tilde{\beta}^{i-1}\!= \!\sum_{p=1}^{[ \frac{i+1}{2} ]} \frac{(i-1)!}{(i-2p+1)!\,(p-1)!^2}\,A^{i-2p+1}R^{p-1} \ .
\raisetag{6pt}
\end{equation}
The values of $\tilde{\delta}^i$ for the first few $i$ are
\begin{equation} \begin{split}
\tilde{\delta}^3 &=  A^2+2R \\
\tilde{\delta}^4 &= A^3+6AR \\
\tilde{\delta}^5 &= A^4+12A^2R+6R^2 \ .
\end{split} \end{equation}

\section{Derivation of the continuum equations}
In this section we shall examine the continuum limit $N \to \infty$, which will allow us to write a simple expression for the generating function $e_2({\bf g})$ of the planar graphs. This will also justify the replacement $Aj \to A, \ R_j \to R$ used in the previous section. Let us consider the identity
\begin{equation} \begin{split}
nh_n & =  \int\dd \mu(\lambda)\, \lambda P_n'(\lambda)P_n(\lambda) \\
{} & =  \int \dd \mu(\lambda)\,P_n'(\lambda)\,[P_{n+1}(\lambda)+R_nP_{n-1}(\lambda)+A_nP_n(\lambda)] \\
{} & =  R_n\int \dd \lambda\, \eu^{-S(\lambda)}P_n'(\lambda)P_{n-1}(\lambda) \\
{} &= R_n\int\dd \lambda\,\eu^{-S(\lambda)}S'(\lambda)P_n(\lambda)P_{n-1}(\lambda) \\
{} &= (1-\sum_{i=3}^k \bar{g}_i\beta^{i-1}_n)\, h_n \, R_n \ ,
\end{split} \end{equation}
where in the last but one equality we have integrated by parts and in the last equality we have used the definition of $\beta_n^i$. Thus we have obtained the first recursion relation
\begin{equation} \label{rec1}
n=(1-\sum_{i=3}^k \bar{g}_i\beta^{i-1}_n)\,R_n \ .
\end{equation}
From this equation we infer in particular that: $R_n({\bf 0})=n$.
We want to find a second recursion relation which relates the coefficients $A_n$ and $R_n$. We observe that:
\begin{equation} \begin{split}
 \int \!  \dd \lambda\, \eu^{-S(\lambda)}\lambda P_n(\lambda)P_{n+1}'(\lambda)   
 &=   \int \!  \dd \lambda \, \eu^{-S(\lambda)}P_n(\lambda)\lambda S'(\lambda)P_{n+1}(\lambda) \\
{} &=  (A_{n}+A_{n+1}-  \sum_{i=3}^k \bar{g}_i \beta^i_{n+1}) \, h_{n+1} \ .
\end{split} \end{equation}
But
\begin{equation} \begin{split}
 \int \!  \dd \lambda\, \eu^{-S(\lambda)}\lambda P_n(\lambda)P_{n+1}'(\lambda)   &=   nA_nh_n+ R_n  \int \!  \dd \lambda \, \eu^{-S(\lambda)}P_{n-1}(\lambda)P_{n+1}'(\lambda) \\
{}&=   n A_n h_n+  R_n  \int \!  \dd \lambda \, \eu^{-S(\lambda)}P_{n-1}(\lambda)S'(\lambda)P_{n+1}(\lambda) \\
{} &=   n A_n h_n-h_{n+1}R_n\sum_{i=3}^k\bar{g}_i \gamma^{i-1}_n \ .
\end{split} 
\raisetag{24pt}
\end{equation}
As a result, the second recursion relation is
\begin{equation} \label{rec2}
(A_n+A_{n+1}-\sum_{i=3}^k\bar{g}_i \beta^i_{n+1})\,R_{n+1}=nA_n-R_{n+1}R_n\sum_{i=3}^k\bar{g}_i\gamma_n^{i-1} \ . 
\end{equation}
Now, we extract the planar case taking the limit $N \to \infty$. Let us introduce the substitutions
\begin{eqnarray}
\frac{n}{N} & \longrightarrow & x \\
\frac{R_n}{N} & \longrightarrow & R(x) \\
\frac{A_n}{\sqrt{N}} & \longrightarrow & A(x)
\end{eqnarray} 
to obtain, taking into account the power of  $N$ contained in  $\bar{g}_i$, the two continuum equations
\begin{eqnarray} \label{sist}
x &= & R(x) \Big( 1-\sum_{i=3}^kg_i\,\tilde{\beta}^{i-1}(x) \Big) \\
 A(x) &=& \sum_{i=3}^k\,g_i\,\tilde{\delta}^i(x) \label{sist2} \ ,
\end{eqnarray}
where $\tilde{\beta}^i(x)$ and $\tilde{\delta}^i(x) $ are expressed in terms of  $A(x)$ and $R(x)$  as given by eqs. (\ref{beta}, \ref{delta}).
One  finds, from eqs. (\ref{rec1}, \ref{rec2}), that the continuous solution $A(x)$, $R(x)$, is related to the coefficients $A_n$ and $R_n$ by
\begin{equation} \begin{split}
\frac{R_n}{N}& =R \left(\frac{n}{N} \right) +O\left( N^{-1} \right) \\
\frac{A_n}{\sqrt{N}}& =A \left(\frac{n}{N} \right) +O\left( N^{-1} \right) \ .
\end{split} \end{equation}
 We are interested only in the continuous solution, indeed, recalling that  $R_n({\bf 0})=n$, we have
\begin{equation}
 \frac{R_n({\bf g})}{R_n({\bf 0})}=\frac{R_n({\bf g})}{N} \left(\frac{n}{N} \right)^{-1} \ ,
\end{equation}
and the function $e_0({\bf g})$ can be rewritten, in the limit $N \to \infty$, as
\begin{equation} \label{inte}
 e_0({\bf g}) = \int_0^1 \dd x \,(1-x) \ln \left(\frac{R(x)}{x} \right) \ .
\end{equation}
We have therefore reduced our problem to the one of obtaining $R(x)$ (or $A(x)$) from the system of continuum equations (\ref{sist}, \ref{sist2}), and to perform the integral in eq. ({\ref{inte}). 
\section{The solution for the cubic vertex}
\label{gamma}
In order to understand the solution for the cubic vertex it is useful to recall the main features of the even vertex case.  
For even vertex $A(x)=0$ and thus the first continuum equation for $R(x)$, 
\begin{equation}
 x=R\, \Big(1-\sum_{p=2}^{k/2}\,g_{2p} \,{2p-1 \choose p} \,R^{p-1} \Big) 
\end{equation}
suffices. The quartic case can be explicitly integrated \cite{Be} to obtain
\begin{equation}
e_0(g_4)=\frac{1}{2} \ln a + \frac{1}{24} (a-1)(a-9)=\sum_{k=1}^{\infty} (3g_4)^k\, \frac{(2k-1)!}{k!\,(k+2)!} 
\end{equation}
with
\begin{equation}
 a= \frac{1-\sqrt{1-12g_4}}{6g_4}=1+\sum_{k=1}^{\infty}\,(3g_4)^k\, 2 \frac{(2k-1)!}{(k+1)!(k-1)!} \ . 
\end{equation}
Recalling  the formula for the topological expansion \eqref{top}, one has the interesting equation
\begin{equation}
 \sum_{\substack{ G \ planar, \ connected,  \\ with \ k \ quartic \ vertices}} \frac{1}{o(A(G))}= 3^k \,\frac{(2k-1)!}{k!\,(k+2)!} 
\end{equation}
that can be checked for $k=1$ and $k=2$ using the contents of figure~\ref{dis2}.
The radius of convergence is $1/12$ and $g_{4c}=1/12$ is the critical point. For $g_4 \to g_{4c}$ one obtains the critical behavior
\begin{equation}
 e_0(g_4) \sim (g_{4c}-g_4)^{\frac{5}{2}} \ ,
\end{equation}
which is particularly interesting for 2D-Gravity.

Coming now to the cubic vertex case, in order to cure the lower unboundedness of the action, one can add a quartic term with negative $g_4$, with the idea to take eventually the limit $g_4 \to 0$. On the other hand it is clear that we can just take $g_4=0$ in all the expressions which turn out to be well defined in that limit. We shall see that the integral \eqref{inte} is well defined for $R(x)$ given by the solution of the continuum equations (\ref{sist}, \ref{sist2}) with $g_4=0$

\begin{equation} \begin{split}
\frac{x}{R}&=1-2g_3A \\
2g_3R &= A-g_3A^2 \ .
\end{split} \end{equation}

In fact, let us introduce the new variable $\sigma=-g_3\,A$ related to  $x$ by
\begin{equation} \label{algeb}
 2g_3^2\, x + \sigma (1+ \sigma) (1 + 2 \sigma)=0 \ .
\end{equation}
The function  $\sigma(x)=\bar{\sigma}(g_3^2\,x)$ is the solution of \eqref{algeb} which vanishes for $x=0$; indeed when $g_3=0$  the potential has no longer odd vertices and then $A(x)=0  \Rightarrow \bar{\sigma}(0)=0$.
Our function  $e_0(g_3)$ may be rewritten, going over from the variable $x$ to the variable $\sigma$ and integrating by parts

\begin{equation} \begin{split}
 e_0(g_3)&= \int_0^1 \dd x\, (1-x) \ln \left( \frac{R(x)}{x} \right) = - \int^{\sigma_1}_0 \dd \sigma \, \frac{ \big( x(\sigma)-1 \big)^2}{1+2\sigma} \\ 
{}&= -\frac{1}{2} \ln(1+2\sigma_1)+\frac{\sigma_1 (2+6\sigma_1+3\sigma_1^2)}{3(1+\sigma_1)(1+2\sigma_1)^2}
\end{split} \end{equation}
where $ \sigma_1=\bar{\sigma}(g_3^2)$ is the solution of
\begin{equation}
 2g_3^2  + \sigma (1+ \sigma) (1 + 2 \sigma)=0 \ ,
\end{equation}
 which vanishes in $g_3=0$. $\sigma_1$ can be expressed as an expansion in powers of $g_3$ using Lagrange theorem,  obtaining
\begin{equation}
\sigma_1=-\frac{1}{4} \sum_{k=1}^{\infty} \left( 8g_3^2 \right)^k \, \frac{\Gamma \big(\frac{1}{2}(3k-1) \big)}{\Gamma(k+1)\, \Gamma \big( \frac{1}{2}(k+1) \big)} \ .
\end{equation}
Except for some factors, due to different definitions, our results coincide with those of Br\'ezin \emph{et al.}~\cite{Br}.  The power expansion series for the planar generating function is
 \begin{equation}
 e_0(g_3)=\frac{1}{2} \sum_{k=1}^{\infty} (8g_3^2 )^k  \,  \, \frac{\Gamma (3k/2)}{\Gamma(k+3)\, \Gamma (k/2+1)} 
\end{equation}
and, recalling the topological expansion for $e_0(g_3)$, we reach the formula
\begin{equation}
 \sum_{\substack{ G \ planar, \ connected, \\  with \ 2k \ cubic \ vertices}} \frac{1}{o(A(G))}=  8^k \frac{1}{2} \,  \frac{\Gamma (3k/2)}{\Gamma(k+3)\,\, \Gamma (k/2+1)} \ .
\end{equation}
Such formula can be checked for $k=1$ and $k=2$ using the contents of figure~\ref{dis1}. The asymptotic behavior of the expansion coefficient is 
\begin{equation}
  8^k \frac{1}{2} \,  \frac{\Gamma (3k/2)}{\Gamma(k+2)\,\, \Gamma (k/2+1)} \sim \frac{\eu^3}{\sqrt{6 \pi}}(12 \sqrt{3})^k k^{-\frac{7}{2}} \ .
\end{equation}

The radius of convergence of the series is ${1/ \sqrt{12 \sqrt{3}}}$ and $g_{3c}={1/ \sqrt{12 \sqrt{3}}}$ is the critical point. From the asymptotic behavior $k^{-\frac{7}{2}}$ we conclude that the critical exponent remains unchanged from the quartic case. 

In 2D-Gravity, where the continuum surfaces are replaced by poligonalizations, such a result is a check of the independence of the partition function, in the limit of infinite number of vertices, of the kind of poligonalization one chooses to approximate the continuum surfaces \cite{GM}.

\section{Conclusions}
In dealing with matrix models  usually one encounters matrix models with even potential so  the question naturally arises  if there is some obstruction to the odd vertex case. In this paper we have shown that, even if, in the odd vertex case,  the  original partition function is ill  defined, the method of orthogonal polynomials can be often applied in its most naive form, that is ignoring all convergence problems. This is justified by adding a regulating even vertex to  the odd one, and taking eventually the limit for its coupling constant  going to zero. We have extended the orthogonal polynomial method to any combination of odd and even vertices, writing the two needed continuum equation. The explicit application to the cubic vertex case has been given, recovering the result of Br\'ezin \emph{et al.} \cite{Br}. An explicit integration of  3+4 or 5 vertex case appears feasible along these lines and would be a useful  check of the universality of the critical behavior.

The general setting explained here  can be readily developed, in the planar case,  also for two-matrix models with coupling in the form of the Itzykson-Zuber formula \cite{IZ,M1}, the cubic case being already solved in \cite{BK}.
Further extension can be developed in  higher genus cases {\em e.g.} in the cubic case for the torus topology.         
\section*{Acknowledgements}
{\em I am grateful to P.Menotti for suggesting this problem and for useful discussions.}


\begin{thebibliography}{bb}
\bibitem{Be} D. Bessis, C. Itzykson, J.B. Zuber, Adv. Appl. Math. {\bf 1} (1980), 109

\bibitem{M1} M.L. Metha, Commun. Math. Phys. {\bf 79} (1981), 327
\bibitem{Br} E. Br\'ezin, C. Itzykson, G. Parisi, J.B. Zuber, Commun. Math. Phys. {\bf 59} (1978), 35 
\bibitem{Be1} D. Bessis, Commun. Math. Phys. {\bf 69} (1979), 147     

\bibitem{BK} D.V. Boulatov, V.A. Kazakov, Phys. Lett. {\bf B126} (1986), 379 
\bibitem{M2} M.L. Metha, {\em Random Matrices}, New York, Academic Press 1967 
\bibitem{GM} P. Ginsparg, G. Moore, {\em Lectures on 2D Gravity and 2D String Theory}, preprint hep-th/9304011 (1993) 
\bibitem{IZ} C. Itzykson, J.B. Zuber, J. Math. Phy. {\bf 21} (1980), 411 
\end{thebibliography}
\end{document}